\def\etal{et al.\ }

\def\oc{$\omega$ Centauri}

\def\c2{\chi ^2}

\documentstyle[aaspp4,psfig]{article}

\begin{document}

\title{The Relation Between Fourier Coefficients and RR Lyrae Magnitudes} 

\author{Ann Berit Saust}
\affil{Princeton University Observatory \\ Princeton, NJ 08544-1001, USA.}

\begin{abstract}
The relation between the Fourier coefficients determined by the light curves of
RR Lyrae variables in the Sculptor dwarf galaxy, the M5 cluster, and the 
unique globular cluster \oc\ is investigated. A couple of recent papers
claim that it is possible to determine both metallicity and absolute
magnitude of RR Lyrae stars from their period and two of the Fourier
coefficients. However, this investigation show that fitting high order Fourier
series to noisy and/or sparse data results in a large scatter in the 
determination of the
absolute magnitudes. Unless we can find a reasonable way of smoothing noisy
data, it will be difficult to use RR Lyrae's as standard candles.
\end{abstract}



\section{Introduction}

It has long been believed that the absolute magnitude of RR Lyrae stars
could be determined by their light curves and, to some still unknown
extent, their metallicity. Fourier decompositions, were introduced
by Simon \& Lee 1981\markcite{si81}, as a means of qualitative
description of light curves of pulsation variables in general, 
and they showed that amplitude ratios and phase
differences provide a useful description of the Hertzsprung progression for
classical Cepheids. In addition,
Simon \& Teays 1982\markcite{si82} showed that the Fourier
decomposition parameters of 70 RR Lyrae field stars are more sensitive 
discriminators of the Bailey type (ab or c) than the traditionally employed
period-amplitude diagram.

Several globular clusters (e.g.\ \oc) contain many RR Lyrae stars, and since 
the variables in a globular cluster probably constitute a much more uniform
sample than the field RR Lyrae's, a considerably smaller scatter is expected
in data from one cluster. Petersen 1984\markcite{pe84} examined Fourier
decompositions of RR Lyrae's in \oc, and found that the scatter in the Fourier
parameters was larger than expected for a uniform sample. Petersen 
1984\markcite{pe84} concluded that the scatter was most likely due to the
large metallicity distribution ([FE/H] ranges from -2.3 to -0.5
(Butler \etal 1978\markcite{bu78}), differences in mass of the horizontal
branch stars, or differences in effective temperature. As Smith 
1995\markcite{sm95} pointed out, \oc\ is so unusual in its range of
chemical composition that it is questionable whether \oc\ is the key to the
absolute magnitude-metallicity relation, too unusual to be a representative
of the RR Lyrae population, or just one more clue to an absolute 
magnitude-metallicity relation relationship which is more complicated than
expected.
See Smith 1995\markcite{sm95} for a review on
RR Lyrae stars in general.

According to two recent papers, Kov\'{a}cs \& Jurcsik 1996\markcite{ko96} (KJ)
and Jurcsik \& Kov\'{a}cs 1996\markcite{ju96} (JK), the metallicity and
absolute magnitude of RR Lyrae ab stars can be described by a simple linear
combination of the period and two Fourier coefficients. 
Kov\'{a}cs \& Jurcsik 1997\markcite{ko97} added multi band observations to their
earlier results to obtain an even smaller error, leading to an estimate of
the relative distance moduli with an accuracy of
$<$ 0.03 mag. However, their 
sample did not include the unique globular cluster \oc, which is included
here. In addition to \oc, this investigation
includes the Sculptor galaxy and the M5 cluster which were also included
in JK and KJ, such that results and errors can be compared directly.

\section{The Observations}

The observations of Sculptor galaxy and \oc\ were made by the OGLE team
during 1992 and 1993 using the 1-m Swope telescope at Las Campanas
Observatory (see Kaluzny \etal 1995\markcite{ka95} and Kaluzny \etal 
1997\markcite{ka97}). The OGLE data is available via anonymous FTP
from astro.princeton.edu, in the directories bp/Sculptor and
bp/Omega\_Cen. All 
The M5 data were made
by N.\ Reid during 1991 and 1992 using the the 60 inch telescope at 
Palomar Observatory (see Reid 1996\markcite{re96}) and is also available 
via anonymous FTP from astro.princeton.edu, in the directory bp/M5. All 
observations used in this paper were done in Johnson V.

\section{Fourier Analysis}

It is customary to decompose the light curve into its Nth order Fourier 
decomposition:
\begin{equation}
M_V ~=~ a_0 ~+~ \sum_{k=1}^{N} a_k sin(2\pi kf) ~+~ b_k cos(2\pi kf)
\end{equation}
where f is the phase given by $f=(JD_{obs} ~-~ JD_0)/P$, $JD_0$ being the mean
epoch and $P$ the period.
Defining $H_k^2 ~=~ a_k^2~+~b_k^2$, $tan(\phi _k )~=~ -a_k/b_k$ this can
be written:
\begin{equation}
M_V ~=~ a_0 ~+~ \sum_{k=1}^{N} H_k sin(2\pi kf ~+~ \phi _k) 
\end{equation}
(see Petersen 1986\markcite{pe86} for more details).

Since the high noise level in especially the OGLE data makes it impossible 
to make a direct high order Fourier fit, all the data was smoothed using
polynomials. Since the same mechanism is believed to be causing the
variability in RR Lyrae's, it would be reasonable that the same smoothing
technique could be applied to all the data. This also has the advantage
that the results are reproducible and can be compared to future observations.

The data were first wrapped according to the period, divided into 9
equal parts which were then smoothed with a
7th order polynomial. The high order of the polynomial ensures
that the curve is smooth in intervals with many data points, but
follows the data points in less crowded intervals. To ensure continuity, 
the actual fitting was done
over a three times larger interval. This combination of windowing and order of
the polynomial gave the smallest $\chi^2$ for all stars, where
\begin{equation}
\chi^2 ~=~ \sum_{i=1}^{N_{obs}}\frac{\mu _{obs,i} ~-~ \mu _{fit,i}}{\sigma _i}
\end{equation}
where $N_{obs}$ is the number of observations, $\mu_{obs,i}$ is the i'th
observed magnitude, $\mu_{fit,i}$ is the i'th calculated magnitude, and
$\sigma _i$ is the noise.
A high (8th) order Fourier series was then
fitted to the smoothed light curve using singular value decomposition as
described in in Press et al.\ (1992\markcite{pr92}). Just as in KJ, the
Fourier parameters refer to a sine decomposition and the phase is chosen as
the closest value to 5.1.
Increasing the order of the Fourier
series did not show any significant changes of the (lower) Fourier 
coefficients. 

\begin{figure}
\centerline{\hbox{\psfig{figure=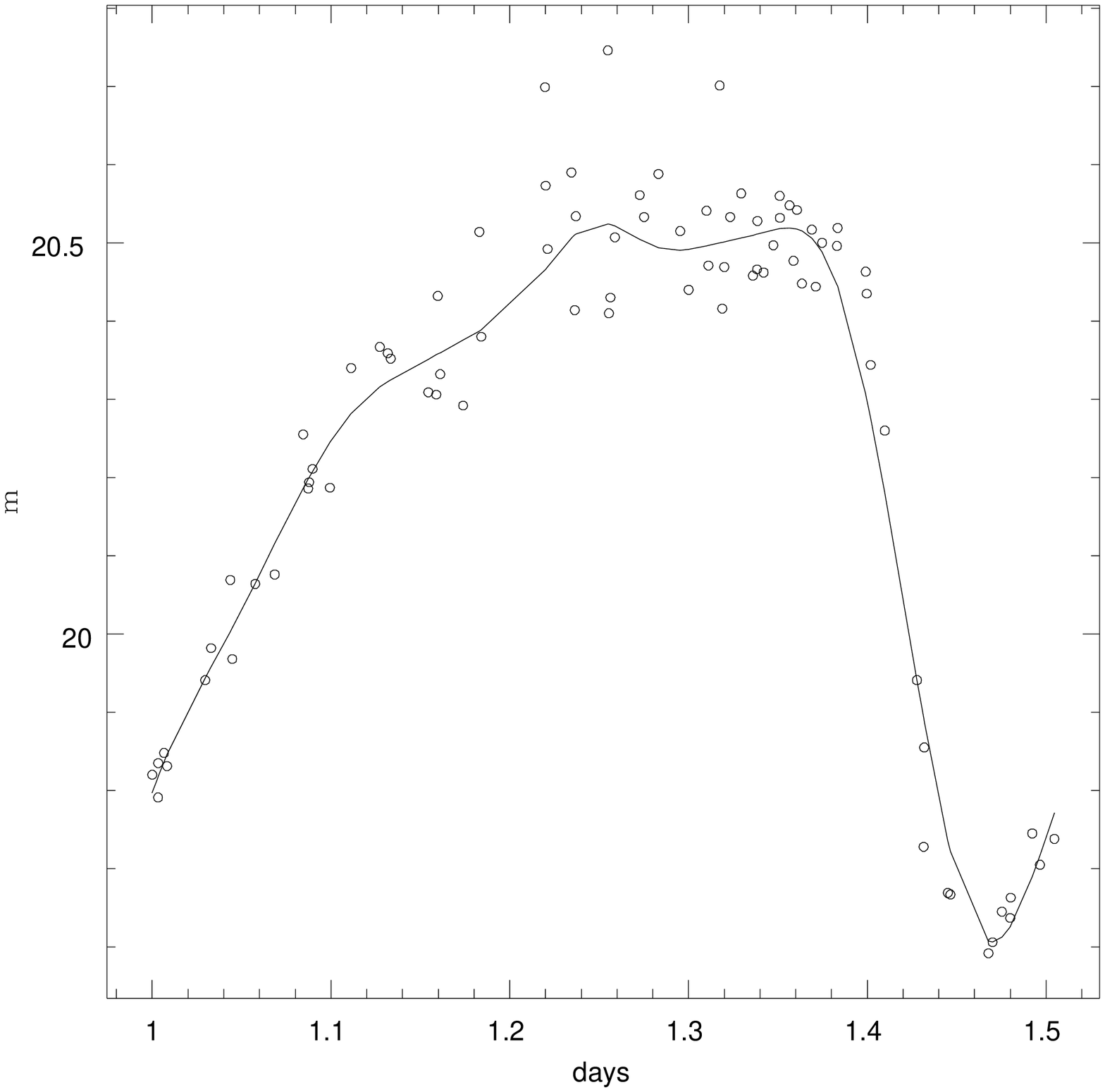,height=2.5in}}}
\caption{Observations and Fourier fit for a star (OGLE-id 37) in the Sculptor 
galaxy.  \label{fit} }
\end{figure}

Fig.\ \ref{fit} shows the observations and the Fourier fit for one star 
from the Sculptor galaxy. As can be seen, a better fit could be obtained
with human aid, e.g.\ using different order of the polynomials in each 
interval, changing the size of the intervals, etc.\ (which is what KJ and JK did
in their papers), but as mentioned above, a general smoothing technique has 
many advantages. No attempts were made to delete or disregard obvious bad 
data points on the individual light curves since this would be impossible
in intervals with sparse data, and one might worry about introducing non
random bias into the data. As a result, the scatter is expected to be fairly
large; however, given sufficient amounts of data the final result should be
similar to KJ's.

JK's main conclusion is that the Fourier
parameter $\phi_{31}$, defined as $\phi_{31} ~=~ \phi_3 ~-~ 3 \cdot \phi_1$
(in general, $\phi_{j1} ~=~ \phi_j ~-~ j \cdot \phi_1$), is the most important
parameter when it comes determining the metallicity and KJ found that the three 
parameters, the period, $P$, $H_1$, and $\phi_{31}$, completely determines 
both magnitude and metallicity.
Their formula for the absolute magnitude is:
\begin{equation}
M_V ~=~ 1.221 ~-~ 1.396P ~-~ 0.477 H_1 ~+~ 0.103 \phi _{31}
\end{equation}
and the metallicity:
\begin{equation}
[Fe/H] ~=~ -5.038 ~-~ 5.394P ~+~ 1.345\phi_{31} \label{metal}
\end{equation}

Following KJ, the observed magnitudes
were fitted as a linear function of $P$, $H_1$, and $\phi_{31}$, again using
singular value decomposition.  To avoid zero point
calculations and problems, the three clusters are treated separately, and 
since RR Lyrae ab stars are both the most abundant in the sample and
the only stars treated by JK and KJ, 
these stars are the only ones included in the following. In addition,
stars showing peculiar behavior are omitted. This includes stars showing 
Blazhko behavior, stars with undetermined periods, a few stars with
periods longer than 1 day, and stars several magnitudes fainter than the
majority of the cluster. This means that 124 
stars in the Sculptor galaxy, 35 M5 stars, and 87 out of 97 \oc\
RR Lyrae ab stars were used.

To estimate how good the fit is, $\chi^2$ is calculated for each fit, where
$\chi^2$ here is defined as:
\begin{equation}
\chi^2 ~=~ \sum_{i=1}^{N_{star}}\frac{m_{obs,i} ~-~ m_{fit,i}}{\sigma _{mean,i}}
\end{equation}
where $N_{star}$ is the number of stars, $m_{obs,i}$ is the i'th
observed mean magnitude, $m_{fit,i}$ is the i'th calculated magnitude, and
$\sigma _{mean,i}$ is the 1 sigma error on the observed mean magnitude.

\begin{figure}
\centerline{\hbox{
  \psfig{figure=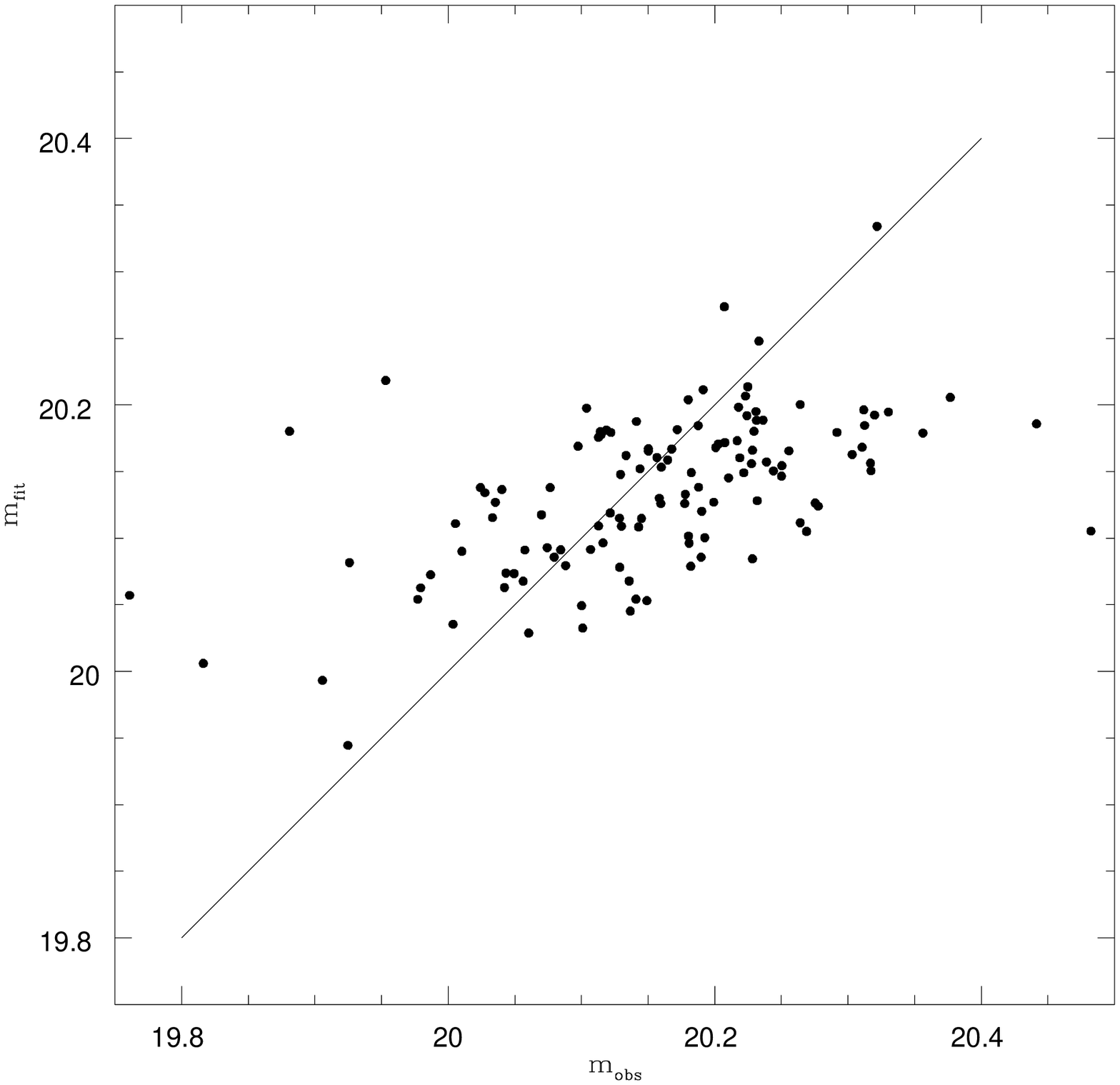,height=2.5in}
  \psfig{figure=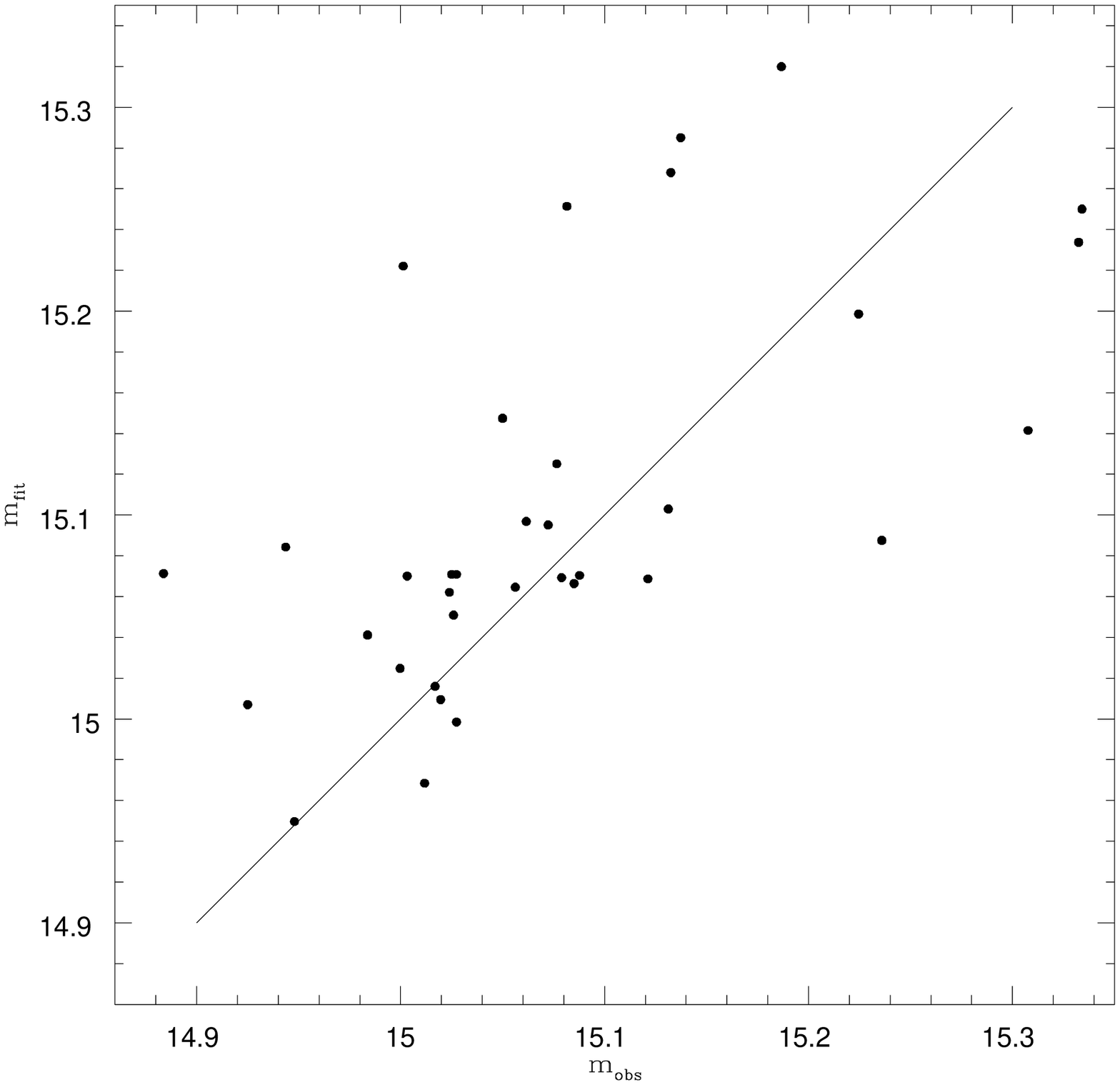,height=2.5in}
  \psfig{figure=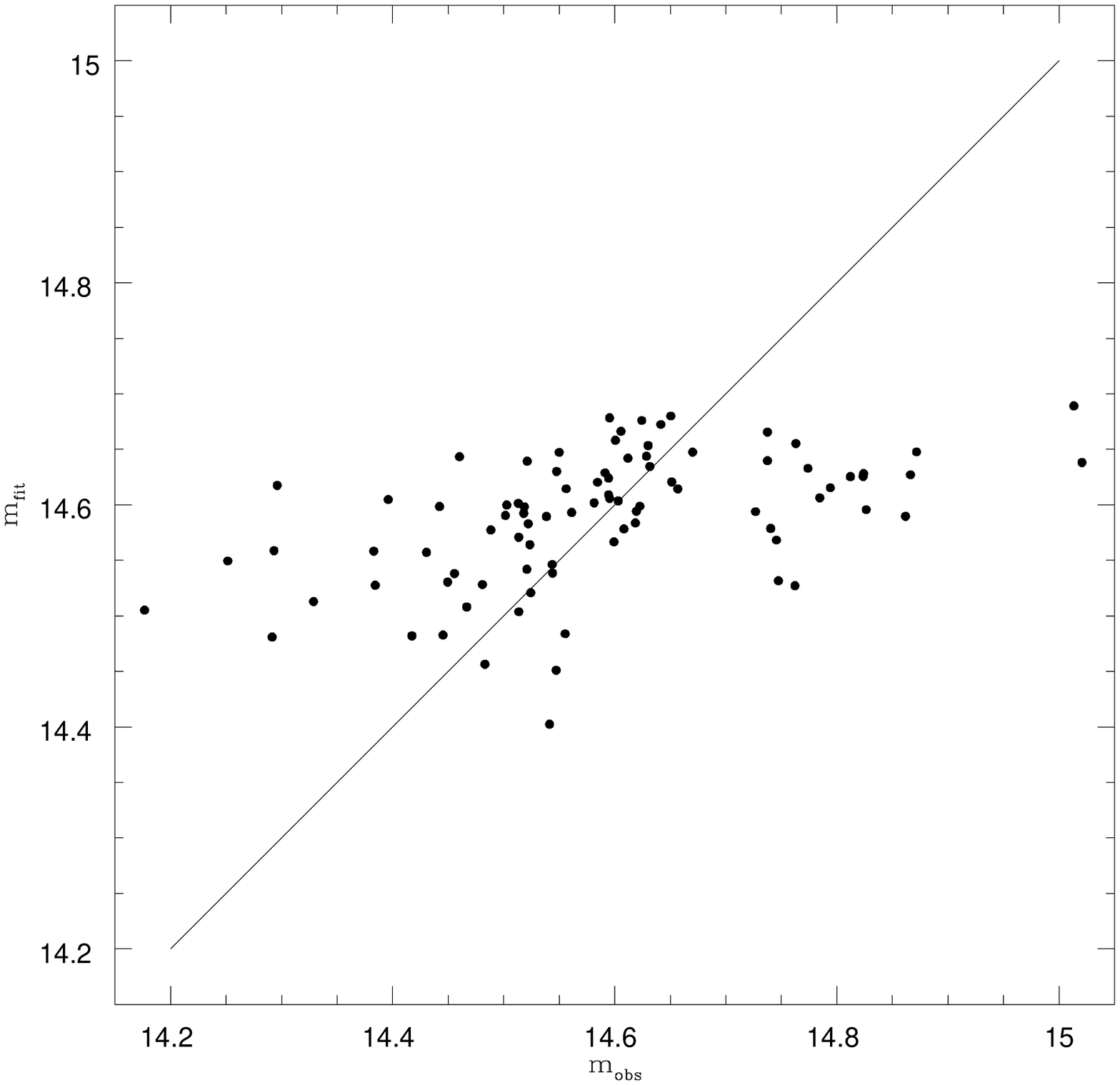,height=2.5in} 
}}
\centerline{(a) \hspace{2.5in} (b) \hspace{2.5in} (c)}
\caption{Observed and calculated magnitudes 
for (a) the Sculptor galaxy, (b) M5, and (b) the \oc\ cluster. The
$m_{fit} ~=~ m_{obs}$ line is added for visibility. \label{fig3} }
\end{figure}

Fitting for the three parameters gives $\chi^2 ~=~ 800$ for
Sculptor, $\chi ^2 ~=~ 77$ for M5, and 
$\chi ^2 ~=~ 750$ for \oc. Fig.\ \ref{fig3} shows the fits which are given by
\begin{equation}
m_S ~=~ 20.45  ~-~ 0.62 P ~+~ 2.96 \cdot 10^{-2} H_1 ~+~ 7.26 \cdot
10^{-3} \phi_{31} 
\end{equation}
\begin{equation}
m_M ~=~ 15.94  ~-~ 1.29 P  ~-~ 7.98 \cdot 10^{-1} H_1 ~+~ 1.47 \cdot 
10^{-3} \phi_{31} 
\end{equation}
\begin{equation}
m_O ~=~ 14.49  ~-~ 1.29 \cdot 10^{-1} P  ~+~ 7.06 \cdot 10^{-1} H_1 ~-~ 
9.33 \cdot 10^{-3} \phi_{31} 
\end{equation}

As can be seen both from the figures and the $\chi^2$'s, the fits are
poor, much poorer than what was found in KJ. Adding additional Fourier
coefficients to the fit gives only a marginal improvement. Since some of the 
Fourier
fits show large $\chi^2$'s, these were omitted in an attempt to reduce
the scatter. However, neither the scatter nor the fit changed significantly,
implying that the main contribution to the scatter is coming from bad data
points in intervals with sparse data (assuming that it is possible to
determine the magnitude as a function of $P$, $H_1$, and $\phi_{31}$ in the
first place). Notice also that there is
almost no dependency on $\phi_{31}$, in contrast to KJ's result, 
and that the dependency on $H_1$ is strongly influenced by the scatter. 
It should also be noted
that while the fits for the Sculptor galaxy and the M5 cluster are somewhat
similar, the fit for \oc\ deviates significantly. The noise level in the
two OGLE data sets are much larger than the noise level in most of
KJ's data, but even in the case
of M5 which has a much lower noise, the $\phi_{31}$ parameter seems to be
negligible.

\begin{figure}
\centerline{\hbox{
  \psfig{figure=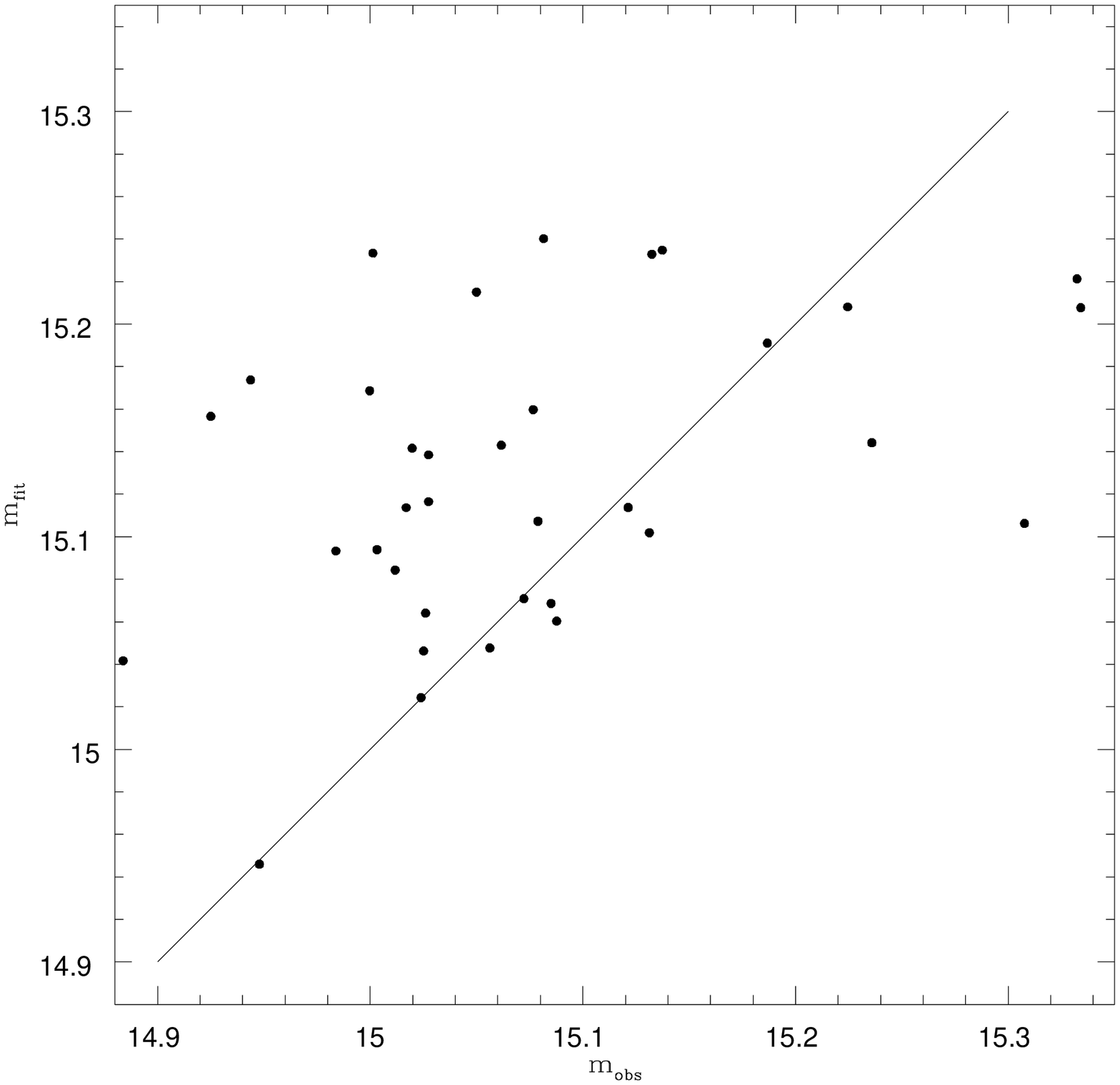,height=2.5in}
}}
\caption{Same as fig.\ 2 above for M5, only no smoothing was applied.
\label{nosmoo} }
\end{figure}

Different smoothing techniques (i.e.\ different windowing or orders of
the fitted polynomials) did not change the fitted parameters
significantly (deviations within 20\%). Since the noise level in the M5
data is low, the M5 data was recalculated without any smoothing. 
Fig.\ \ref{nosmoo} shows the result, $\chi^2$ is 93 here, and the fit is 
given by:
\begin{equation}
m_M ~=~ 15.79  ~-~ 1.09 P  ~-~ 1.16 \cdot 10^{-1} H_1 ~-~ 6.14 \cdot 
10^{-3} \phi_{31} 
\end{equation}
Again, the dependence on the $\phi_{31}$ parameter is small.

\begin{figure}
\centerline{\hbox{
  \psfig{figure=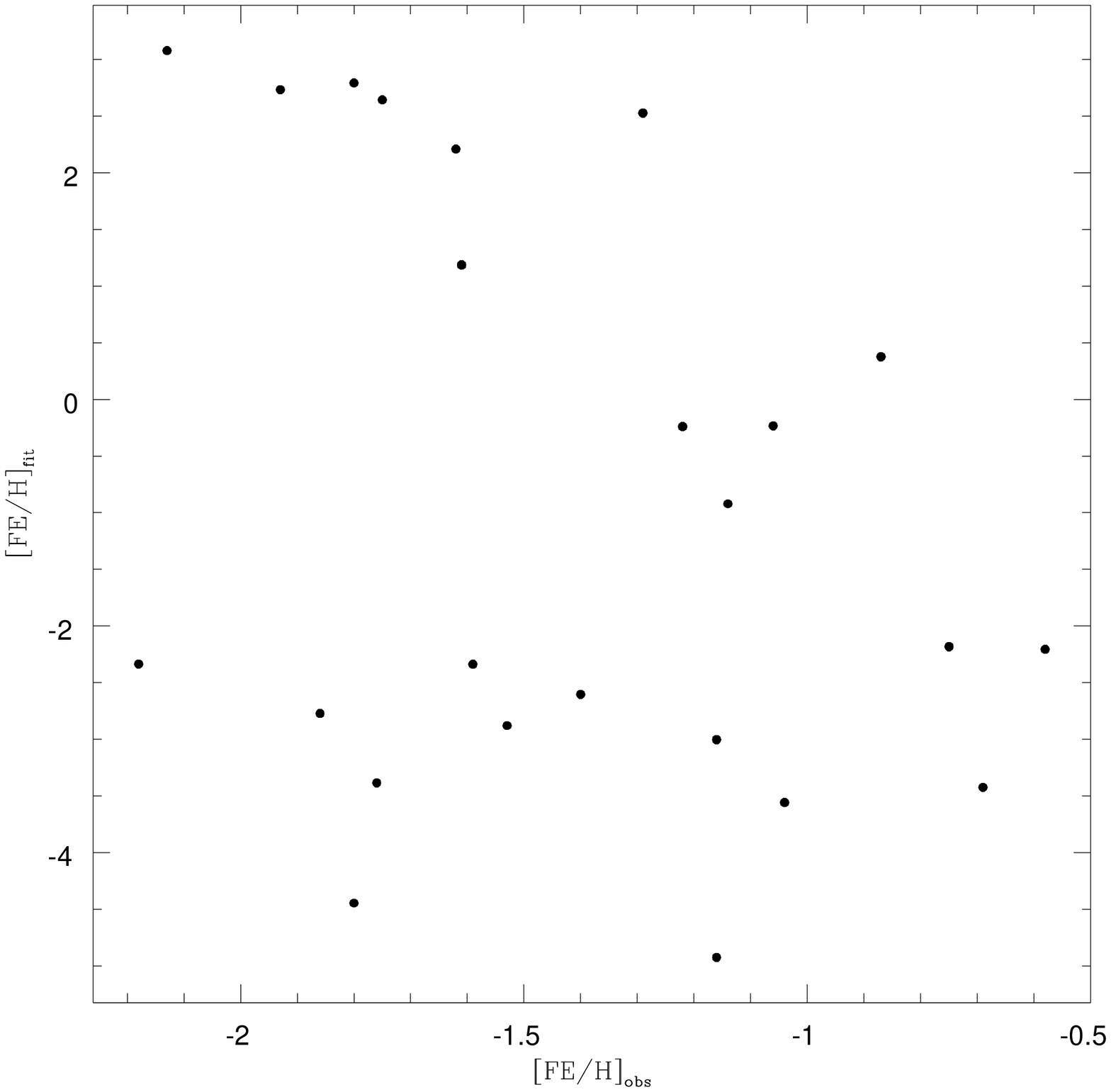,height=2.5in}
}}
\caption{Observed and calculated [Fe/H] for \oc. [Fe/H] was calculated according
to the formula given by Jurcsik \& Kov\'{a}cs 1996. 
\label{fig6} }
\end{figure}

Since the metallicity distribution in \oc\ is larger than in any other cluster, 
it would be interesting to do the same exercise for the metallicity. 
Unfortunately, there is 
not enough data available in the literature, and the metallicity of the \oc\
stars were instead calculated using the result from
JK (equation \ref{metal} above) and
compared to the observed [Fe/H] values from Butler \etal 1978\markcite{bu78}.
Notice that there are only 18 stars which are common for the two data sets, 
The result is shown in fig.\ \ref{fig6} and, as can be seen, resembles a
scatter plot.

\section{Discussion}

Although the observations in this paper are noisy, the deviations from earlier
results are troublesome in both the case of Sculptor and M5, where it was
impossible to reproduce the low scatter found by KJ, and in the case of \oc\ 
where none of the earlier fits seems to fit the observations.

Since KJ also included Sculptor and M5 in
their analysis, their fits should show scatter similar to what was found above.
It seems that the deviations for Sculptor and M5 are mostly
caused by the different smoothing techniques. However, any general smoothing
using polynomials all give results similar to the above, and looking at the
individual light curves, any smoothing (general or done with human aid)
is a non trivial task due to the noisy and/or sparse data. In addition,
the fits for both Sculptor and \oc\ (fig.\ \ref{fig3}) shows evidence of 
extinction.

Due to the unique metallicity distribution in \oc\, this cluster can be
regarded as either the best cluster to determine relations between
absolute magnitude and metallicity or a non representative example of
the RR Lyrae population. This investigation seems to support the latter;
however, the concerns over the smoothing makes it impossible to
make a conclusion.

Since noisy and sparse data is common place, a general smoothing technique would
be desirable, and it seems that if we want to use RR Lyrae's as 
standard candles, some form of general consensus on the smoothing has to be 
reached.  It is by no means certain that the smoothing applied here is
the best available, and it would not be surprising if the scatter is reduced
by a different way of smoothing. However, based on the above, it seems unlikely
that any general smoothing would reduce the scatter to the level found by
KJ. Since the above results strongly suggest that the main part of the
scatter is due to bad data points in intervals with sparse data,
a better sampling of data would obviously improve the results considerably.

\acknowledgments {\em Acknowledgments:} 
I would like to thank the OGLE team and N.\ Reid for obtaining the data
and making it easily available via anonymous FTP. I would also like
to thank J. Jurcsik and B.\ Paczy\'{n}ski for discussions and helpful
suggestions.
This work was supported by NSF grants AST-9530478 and AST-9528096.


\newpage

\begin{references}

\reference{bu78} Butler, D., Dickens, R.J., Epps, E.: 1978, ApJ 225, 148

\reference{ju96} Jurcsik, J., Kov\'{a}cs, G.: 1996, A\&A 312, 111 (JK)

\reference{ka95} Kaluzny,J., Kubiak, M., Szyma\'{n}sky, M., Udalski, A., 
   Krzemin\'{s}ki, W., Mateo, M.: 1995, A\&AS 112,407

\reference{ka97} Kaluzny,J., Kubiak, M., Szyma\'{n}sky, M., Udalski, A., 
   Krzemin\'{s}ki, W., Mateo, M.: 1997, in preparation

\reference{ko96} Kov\'{a}cs, G., Jurcsik, J.: 1996, ApJ 466, L17

\reference{ko97} Kov\'{a}cs, G., Jurcsik, J.: 1996, A\&A 322, 218

\reference{pr92} Press, W.\ H., Flannery, B.\ P., Teukolsky, S.\ A., Vetterling,
   W.\ T.: 1992, ``Numerical Recipes'' (Cambridge University Press)

\reference{re96} Reid, N.: 1996, MNRAS 278, 367

\reference{pe84} Petersen, J.O.: 1984, A\&A 139, 496

\reference{pe86} Petersen, J.O.: 1986, A\&A 170, 59

\reference{si81} Simon, N.R., Lee, A.S.: 1981, ApJ 248, 291

\reference{si82} Simon, N.R., Teays, T.J.: 1982, ApJ 261, 586

\reference{sm95} Smith, H.A.: 1995, ``RR Lyrae Stars'' (Cambridge University Press)

\end{references}
\end{document}